\documentclass[12pt]{article}
\usepackage{amsmath,amsfonts,amssymb,graphicx,epsfig}
\def\ba{\begin{eqnarray}}
\def\ea{\end{eqnarray}}

\def\nn{\nonumber}
\usepackage{graphicx}

\begin{document}
\title{\Huge\bf On Dirac-like Monopoles in a Lorentz- and CPT-violating Electrodynamics}
\author{{{N.M. Barraz Jr.}$\,^a$, {J.M. Fonseca}$\,^a$, {W.A. Moura-Melo}$^{\,a}$\thanks{E-mail: winder@cbpf.br} , and {J.A. Helay\"el-Neto}$^{\,b,c}$\thanks{E-mail: helayel@cbpf.br}}\\ \\
\it $^a$ \small \it Departamento de F\'{\i}sica, Universidade Federal
de Vi\c{c}osa\\ \small \it 36570-000, Vi\c{c}osa, Minas
Gerais, Brazil.\\
\small
\it $^b$ Centro Brasileiro de Pesquisas F\'{\i}sicas\\ \small \it Rua Xavier
Sigaud,
150, Urca, 22290-180, Rio de Janeiro, Brazil\\
\small \it $^c$ Grupo de F\'{\i}sica Te\'orica Jos\'e Leite Lopes\\ \small \it Petr\'opolis, RJ, Brazil}

 \date{}
\maketitle
\begin{abstract}
We study magnetic monopoles in a Lorentz- and CPT-odd electrodynamical framework in (3+1) dimensions. This is the standard Maxwell model
extended by means of a Chern-Simons-like term, $b_\mu\tilde{F}^{\mu\nu}A_\nu$ ($b_\mu$ constant), which respects gauge invariance but violates both Lorentz and CPT symmetries (as a consequence, duality is also lost). Our main interest concerns the analysis of the model
in the presence of Dirac monopoles, so that the Bianchi identity no
longer holds, which naively yields the non-conservation of electric charge. Since gauge symmetry is respected, the issue of charge conservation is more involved. Actually, the inconsistency may be circumvented, if we assume that the appearance of a monopole induces an extra electric current. The reduction of the model to (2+1) dimensions in the presence of both the magnetic sources and Lorentz-violating terms is presented. There, a quantization condition involving the scalar remnant of $b_\mu$, say, the mass parameter, is obtained. We also point out that the breaking of duality may be associated with an asymmetry between electric and magnetic sources in this background, so that the electromagnetic force experienced by a magnetic pole is supplemented by an extra term proportional to $b_\mu$, whenever compared to the one acting on an electric charge.
\end{abstract}

\newpage

\section{Introduction and Motivation}
\indent Since the seminal paper by Dirac on magnetic monopoles\cite{Dirac1931} this has become a recurrent issue in the Physics literature. Among other features, such a work put forward an example illustrating the relation between field topology and Physics, say, how the non-trivial topology of the vector potential yields the physical (however, not yet observed!) magnetic
pole. Nowadays, topics involving topology and physical systems are observed in several
branches of Physics and related sciences. For instance, it is now widely
recognized the role played by topological excitations in the properties of a
number of realistic systems, namely, phase transition and related aspects. Even
though magnetic monopoles have not been observed by now, they  have triggered a
great deal of efforts and their presence (or absence) has been investigated in
the framework of several models, both Abelian and non-Abelian versions, defined
in a number of space-time dimensions.\\

On the other hand, it is well-known that symmetries are some of the keystones
of modern physical theories. Particularly, in High Energy Physics, a ``good
model'' is expected to respect Lorentz- and CPT-symmetries, besides gauge and
other eventual invariances. However, models which do not share such a property are not absent in the
literature. As far as we know, the first local theory proposed with such feature was pointed forward in the work of Ref.\cite{CFJ} concerning a possible
mass generation mechanism for photon mass in (3+1) dimensions which
keeps gauge but violates Lorentz and CPT symmetries. More recently, such a kind of terms
were reconsidered, for example, in connection with birefringence effects observed in astrophysics \cite{birefringence} and also in connection with radiatively
induced Chern-Simons-like term in (3+1)-dimensional Electrodynamics,
starting off from a Lorentz- and CPT-violating model\cite{JackiwKosteleck'y}.
Ever since, a number of articles has been devoted to study models containing
such sort of terms and several classical and quantum results were obtained (see,
for instance, Refs.\cite{ColladayKosteleck'y,Boldoetal,Kosteleck'y2004}, and
related references cited therein). Namely, the issue of 't Hooft-Polyakov monopoles was studied in Ref. \cite{BSHN}, where a $SO(3)$ non-Abelian model violating those symmetries was considered. Indeed, the authors have shown that whenever $b_\mu$ is pure space-like then static and topologically stable solutions are supported. Besides a number of peculiarities on
the vector boson excitation spectrum after the internal symmetry is
spontaneously broken, the role of the background vector that realizes the Lorentz-symmetry violation plays in triggering the monopole formation becomes clear. In the framework of an extended supersymmetry with
Lorentz-symmetry violating term, the monopole solution would induce a non-trivial contribution to the central charge which depends on the
background vector and, then, the bounds on this vector would dictate possible bounds on the supersymmetry central charge triggered by the monopole solution.\\

Here, we would like to carry out a discussion on the presence of
magnetic monopoles in this framework, say, at the Abelian level of this Electrodynamics (sometimes referred as Carroll-Field-Jackiw model). Actually, it has been claimed that Dirac monopole cannot be suitably introduced in this framework\cite{Boldoetal,BSHN}, once their presence implies in charge non-conservation. Among others, we would like to show that this inconsistency may be circumvented provided that the appearance of the monopole is followed by an induced (extra) electric current, analogously to what happens in the (2+1)-dimensional Maxwell-Chern-Simons (MCS) electrodynamics\cite{HT}. In order to explore the similarities and differences between the present and the MCS theories in the presence of magnetic sources we carry out a dimensional reduction of the first to (2+1) dimensions. There, one of the remaining models is MCS with both Lorentz-violating and magnetic pole. We also pay attention to the energy-momentum and electromagnetic force in (3+1) dimensions. Once duality is lost, with $b_\mu\neq0$, we notice an asymmetry between the electric and magnetic sources, which implies in an ambiguity whenever obtaining the expression for the electromagnetic force: it seems that there is an extra term associated to the Lorentz- and CPT-violating background acting only on the magnetic sources. This term is proportional to $\int\,A_\mu k^\mu \, dV$, where $A_\mu$ is the potential and $k^\mu$ the magnetic current. Then, whether the extra force is non-vanishing strongly depends on the effects of the background on these quantities, so that the explicit Lorentz invariant character of that product, $A_\mu {k}^\mu$, could be broken. The trouble lies on the fact that a decision on the vanishing of such term is not straightforward in this framework, since we do not exactly know how it is affected by the background.\\

Our paper is outlined as follows: in Section 2, we review the model and some of
its basic properties. Section 3 is devoted to discuss how magnetic sources may be suitably introduced in this model, by allowing the appearance of an induced (extra) electric current. In Section 4, we perform a dimensional reduction of the model in the presence of electric and magnetic sources to (2+1) dimensions and pay attention to some related issues. Section 5 is devoted to investigate a possible extra force experienced by magnetic sources in this framework. Finally, we close our work by pointing
out our Conclusions and Prospects for future investigation.

\section{The model and some basic aspects}
We consider the following Lagrangian model, defined in (3+1)
dimensions\footnote{Our conventions read: $\mu,\nu, {\rm etc.}\,=0,1,2,3, \,
{\rm diag}\,(\eta_{\mu\nu})=(+,-,-,-)$ and $\epsilon^{0123}=-\epsilon_{0123}=+1$,
etc; $\tilde{F}^{\mu\nu}=\frac12 \epsilon^{\mu\nu\alpha\beta}
F_{\alpha\beta}$. We also use natural units, then, $\hbar=c=1$, etc,
except in some specific points where their explicit presence is important.}:
\begin{equation}
{\cal L}_{\rm total}={\cal L}_{\rm EM}+{\cal L}_{\rm sources}\nn\,,
\end{equation}
where the electromagnetic field sector reads
\begin{equation}
{\cal L}_{\rm EM}=-\frac{1}{4} F_{\mu\nu}F^{\mu\nu}-\frac12 b_{\alpha}
A_{\beta}\tilde{F}^{\alpha\beta}\,,\label{LEMp}
\end{equation}
while ${\cal L}_{\rm sources}$ accounts for the (matter) sources. The second term in eq. (\ref{LEMp}), which resembles us the
Chern-Simons one in (2+1) dimensions will be better explained below.

Indeed, in the presence of electric source, ${\cal L}_{\rm sources}=A^\mu J_\mu$, the dynamical and geometrical equations read like follows:
\ba
& & \partial_{\mu}F^{\mu\nu}=J^\nu + b_\mu \tilde{F}^{\mu\nu}\label{dyneq}\\
& & \partial_\mu\tilde{F}^{\mu\nu}=0\label{bianchi}\,,
\ea
where the latter manifests the absence of magnetic sources, $k_\mu=0$. This set of equations also states us that, even in the absence of electric and/or magnetic
4-currents, the electromagnetic fields are sources for themselves. This means that photon-photon coupling is a possible interaction in this framework, similarly to what happens in the (2+1)-dimensional counterpart provided
by the Chern-Simons action. In addition, it is also known that the present model
displays two distinct excitations in its spectrum, one massive and the another
massless \cite{CFJ}.\\

In turn, the energy-momentum tensor for the electromagnetic field takes the
form \cite{CFJ} (see also Refs.\cite{itin,hehlitin} for an alternative approach
to this quantity):
\begin{equation}
\Theta^{\mu\nu}=-F^{\mu\alpha}F^\nu_\alpha +\frac{\eta^{\mu\nu}}{4}
F^{\alpha\beta}F_{\alpha\beta} +\frac12 b^\nu \tilde{F}^{\mu\alpha}
A_\alpha\,,\label{energymomentump}
\end{equation}
where the latter contribution comes about due to the Chern-Simons-like term.
Indeed, if we demand that such
a term does not breaks gauge invariance, then we should set up $\partial_\mu\,
b_\nu=0$ (in a Minkowskian space-time), implying that $b_\mu$ is a
constant vector, picking up a preferred direction in a given coordinate frame. This anisotropy of the space-time yields violation of Lorentz- and CPT-symmetries \cite{CFJ}. As a consequence, it follows that this tensor no longer presents a null trace, $\Theta^\mu_{\,\;\mu}\neq0$.\\

Another point we should stress concerns the behavior of the energy-momentum tensor (\ref{energymomentump}) under gauge transformations, $A_\mu\,\rightarrow\,A_\mu+e\partial_\mu\,\Lambda$. On a first
glance, this quantity is not gauge invariant, since it explicitly depends on
the vector potential. Actually, although the density of energy and momenta
depends on the gauge, it has been shown that the total energy-momentum is gauge
invariant, provided that potential and fields vanish at infinite \cite{CFJ}.  Moreover, in the absence of sources $\partial_\mu \Theta^{\mu\nu}=0$, while in the presence of electric ones the Lorentz force  immediately follows:
\begin{equation}
\frac{d}{d\tau}P^\mu|_q=q\,u_\alpha\,F^{\mu\alpha}\,,\label{FLorentzq}
\end{equation}
where $q$ is the electric charge and $u^\mu=\gamma(1,\vec{v})$ its 4-velocity.\\

Another interesting aspect to stress is that the electromagnetic duality does not hold in this scenario. Actually, notice that eqs. (\ref{dyneq})-(\ref{bianchi}) transform into each other, with $J_\mu=0$, by means of:
\ba
\left\{
\begin{array}[1]{l}
F_{\mu\nu}\;\longrightarrow\; \tilde{F}_{\mu\nu} +b^\alpha(\eta_{\mu\alpha} A_\nu -\eta_{\mu\nu} A_\alpha)\label{dual1}\,,\\
\tilde{F}_{\mu\nu}\;\longrightarrow\; F_{\mu\nu} +\epsilon_{\mu\nu\alpha\beta} b^\alpha A^\beta \label{dual2}\,.\end{array} \right.
\ea
However, the original eqs. (\ref{dyneq})-(\ref{bianchi}) are no longer recovered if we take the dual of the transformations above. In other words, the proposed duality holds once, but it fails whenever applied twice and so forth. Alternatively, we should notice that the action obtained from Lagrangian (\ref{LEMp}) cannot be made invariant under these (or any other) duality transformations. We would like to return to this point later, when we shall discuss on a possible extra force experienced by the magnetic sources (see section 5, for details).\\

\section{Magnetic sources and the induced current}
To introduce a Dirac monopole in the present model, we proceed as usually: the Bianchi identity (\ref{bianchi}) is now broken by the presence of a magnetic current, $k_\mu$, like it follows below:
\ba
\partial_\mu\tilde{F}^{\mu\nu}=k^\nu\neq0\label{bianchibroken}\,,
\ea

A naive inspection of eqs.(\ref{dyneq}) and (\ref{bianchibroken}) clearly shows the apparent contradiction of the model in the presence of magnetic sources: If we take the divergent in (\ref{dyneq}), we get:
\ba
\partial_\nu \partial_{\mu}F^{\mu\nu}=\partial_\nu J^\nu + b_\mu \partial_\nu\tilde{F}^{\mu\nu}=\partial_\nu J^\nu -b_\mu k^\mu=0\,,\label{chargenonconservation}
\ea
what implies that $J_\mu$ is not conserved at this level, say:
\ba
\partial_\mu{J}^\mu=b_\mu{k}^\mu\neq0 \,.\label{divJneq0}
\ea
This result has been used to prevent the introduction of Dirac-like monopoles in this model, since its presence spoils the (electric) charge conservation \cite{Boldoetal,BSHN}. This is, however, a subtle point once gauge symmetry is respected. Actually, a similar subtitle occurs in the (2+1)-dimensional Maxwell-Chern-Simons model. There, Henneaux and Teitelboim solved the problem by admitting that the presence of a magnetic pole naturally induces the appearance of an extra electric current, which comes about in connection with the topology of the model and it is proportional to $mg$ (Chern-Simons mass times monopole strength). Among other results, they showed that $m$ appears quantized in units of $N\pi\hbar c/g^2$ (for details see Ref. \cite{HT}; see also \cite{Pisarski,prd652002}).\\

In what follows, we would like to show that a similar procedure holds here and renders the presence of Dirac monopole in this framework. For that, let us return to expressions (\ref{dyneq},\ref{chargenonconservation}) and note that, if we introduce in (\ref{dyneq}) an extra induced current, $j^\mu_{\rm ind}$, given by:
\ba
j^\mu_{\rm ind}(z)= b_\alpha\tilde{F}^{\alpha\mu}\,,\label{jind}
\ea
then, the total current, $J^\mu_{\rm total}=J^\mu+j^\mu_{\rm ind}$, is now conserved, $\partial_\mu J^\mu_{\rm total}=0$. Above, $b_\alpha\tilde{F}^{\alpha\mu}$ is valued along the induced current world-line, $z^\mu$. A $\theta(x^0)$ (step function) may also be introduced in the current above for ensuring causality: $j^\mu_{\rm ind}$ is induced only after the magnetic sources appear. Alternatively,  $j^\mu_{\rm ind}$ may be non-locally written in terms of the product $b_\alpha k^\alpha$ by suitably integration of (\ref{divJneq0}). It is instructive to consider the simplest case of a static poin-like magnetic monopole, $k^\mu=(g\delta^3(\vec{r}); \vec{0})$ giving $\vec{B}=g\vec{r}/|\vec{r}|^3$. In this case, we get $j^\mu_{\rm in}=(\-\vec{b}\cdot\vec{B}; b_0\vec{B})$ which is well-behaved but yields a `blowing up charge' whenever integrated over all space, say:
\ba
Q_{\rm ind}=-g\vec{b}\cdot\int d^3r \frac{\hat{r}}{r^2}= -4\pi g\vec{b}\cdot\hat{r}\int dr\,.
\ea
Similarly, the total induced current reads, $\vec{I}_{\rm ind}=4\pi b_0 \int dr \hat{r}$. For extracting finite results from quantities above we must impose a suitable spatial cutoff to these integrals or, alternatively, take $b_\mu$ restricted to a finite region of space-time. On the other hand, if we slice the space the projection of $Q_{\rm ind}$ on the $xy$ plane would give $q^{xy}_{\rm ind}=-2\pi g|\vec{b}|$, which would imply the quantization of the $|\vec{b}|$-parameter, by invoking Dirac condition, $qg=2\pi\hbar cN$. This is, however, an artificial procedure. To correctly implement such a `slice', we should systematically reduce the model to (2+1) dimensions, where a suitable quantization condition involves the remnant of $b_\mu$.\\

\section{Dimensional reduction and related subjects}
In order to relate our previous analysis to their (2+1)-dimensional counterpart, we propose to carry out a dimensional reduction of the present Electrodynamics, with both electric and magnetic currents, to (2+1) dimensions. In the absence of magnetic sources, its Lagrangian has been dimensionally reduced in the works of Refs.\cite{belichetal,dimred}. In our case, however, such a procedure can only be performed at the level of the eqs. of motion (on-shell), once a (classical) Lagrangian description based upon a unique potential, $A_\mu$, bearing a singular structure is not possible.\\

Let us write down again our starting equations:
\ba
\left\{
\begin{array}[1]{l}
 \partial_{\mu}F^{\mu\nu}=J^\nu +j^\nu_{\rm ind}+ b_\mu \tilde{F}^{\mu\nu}\\
 \partial_\mu\tilde{F}^{\mu\nu}=k^\nu\,,\label{seteqprered} \end{array}\right.
\ea
where we have also introduced $j^\nu_{\rm ind}$ given by (\ref{jind}). Our reduction scheme is based upon the following ansatz over any 4-vector: i) its $0,1,2$ components are kept untouched and join each other to form the remaining 3-vector; ii) the 3rd component is identified with a Lorentz scalar in (2+1) dimensions; iii) all the fields no longer depends on the 3rd spatial derivative, $\partial^3 (\rm anything)\equiv0$. Explicitly, we then have:

\ba \left\{
\begin{array}[1]{l}
 A^\mu\,\rightarrow\;(A^{\hat{\mu}}; \phi)\,, \\
        b^\mu\;\rightarrow\;(b^{\hat{\mu}}; m)\,, \\
	j^\mu\;\rightarrow\;(j^{\hat{\mu}}; j)\,,\\
	 k^\mu\;\rightarrow\;(k^{\hat{\mu}}; \chi)\,, \label{DR}
\end{array} \right.
\ea
where $\hat{\mu},\hat{\nu}, \,{\rm etc}=0,1,2$ with the metric signs ${\rm diag} (\eta^{\hat{\mu}\hat{\nu}})=(+,-,-)$. For distinguishing between the two electric currents we take $(J^{\hat{\mu}}; J^3\equiv\lambda)$ and $(j^{\hat{\mu}}_{\rm ind}; j^3_{\rm ind}\equiv \lambda_{\rm ind})$ in (2+1) dimensions. Now, carrying out the reduction of the set of eqs. (\ref{seteqprered}), we are left with:
\ba
& & \left\{
\begin{array}[1]{l}
\partial_{\hat{\mu}}F^{\hat{\mu}\hat{\nu}}=J^{\hat{\nu}} +j^{\hat{\nu}}_{\rm ind} -m\tilde{F}^{\hat{\nu}} +b_{\hat{\alpha}}\tilde{G}^{\hat{\alpha}\hat{\nu}}\,\\
\partial_{\hat{\mu}}\tilde{F}^{\hat{\mu}}=k^3=\chi\,,       \label{2+1model}
\end{array} \right.\ea
\ba
& & \left\{
 \begin{array}[1]{l}
\partial_{\hat{\mu}}G^{\hat{\mu}}=\lambda+\lambda_{\rm ind} + b_{\hat{\alpha}}\tilde{F}^{\hat{\alpha}}\,\\
\partial_{\hat{\mu}}\tilde{G}^{\hat{\mu}\hat{\nu}}=k^{\hat{\nu}}\,,
    \label{2+1model-escalar}
\end{array} \right.
\ea
where the new field-strengths are defined as $F^{\hat{\mu}\hat{\nu}}=\partial^{\hat{\mu}}A^{\hat{\nu}}- \partial^{\hat{\nu}}A^{\hat{\mu}}$ and $G^{\hat{\mu}}=\partial^{\hat{\mu}} \phi$, while their duals read $\tilde{F}^{\hat{\mu}}=\frac12 \epsilon^{\hat{\mu}\hat{\nu}\hat{\kappa}} F_{\hat{\nu}\hat{\kappa}}$ and $\tilde{G}^{\hat{\mu}\hat{\nu}}= \epsilon^{\hat{\mu}\hat{\nu}\hat{\kappa}} G_{\hat{\kappa}}$. The (2+1)-dimensional electric and (pseudoscalar) magnetic fields are stored in $\tilde{F}^{\hat{\mu}}=(-B; -\epsilon^{\hat{i}\hat{j}}E^{\hat{j}})$.\\

The first set of equations define a genuine (2+1)-dimensional MCS electrodynamics with both Lorentz-violating and magnetic source terms. Under the action of CPT, it may be even or odd depending whether or not $b^{\hat{\mu}}$ is  spacelike\cite{belichetal,dimred}. Concerning the gauge symmetry, it is preserved under our dimensional reduction scheme, so that this model is invariant under $A^{\hat{\mu}}\; \rightarrow\; A^{\hat{\mu}}+\partial^{\hat{\mu}}\Lambda$. The second set defines a Scalar Electrodynamics and is coupled to the former by means of the Lorentz-breaking parameter. Furthermore, the three on-shell degrees of freedom (df) associated to the physical excitation in (3+1) dimensions\cite{CFJ} are now stored in the massive $A^{\hat{\mu}}$ (two ones, which are $\pm m$) and one in the massless scalar Klein-Gordon field, $\phi$ (by iterating eqs. above we obtain $\partial_\mu\partial^\mu\phi=0$, in the absence of sources).\\

Now, let us concentrate on the vectorial (2+1)-dimensional Electrodynamics, whose dynamical and geometrical equations correspond to the set (\ref{2+1model}). If we apply $\partial_{\hat{\nu}}$ to the 1st equation of this set, we see that the induced current should satisfy:
\ba
\partial_{\hat{\nu}}j^{\hat{\nu}}_{\rm ind}= b_{\hat{\nu}}k^{\hat{\nu}} -m\chi\,,\nn
\ea
mixing the magnetic-like sources associated to the distinct models. Now, if we consider the simplest case of an instanton-like pole, $\chi=g\delta^2(\vec{x})\delta(t)$, in the vectorial model we are left with the problem studied by Henneaux and Teitelboim\cite{HT} from which emerges the quantization of the Chern-Simons mass according to $m=2\pi\hbar N/g^2$, which is the 3rd-component of the original $b_\mu$-parameter.\\

We may wonder whether a quantization condition may explicitly involve also the components of $b_{\hat{\mu}}$. This seems to be the case if we consider the breaking of Bianchi identity in both, vectorial and scalar electrodynamics, eqs. (\ref{2+1model}) and (\ref{2+1model-escalar}), respectively. In this case, we should note that the models appear to be coupled. A conclusion upon this question is more involved and has been worked out \cite{workinprogress}.\\

\section{Electromagnetic force and the asymmetry between electric and magnetic sources}

We have seen, in Section 2, that the energy-momentum tensor yields the usual Lorentz force acting on electric sources, so that no energy or momentum related to the Lorentz-violation is transferred from the fields to the sources and vice-versa. Whenever $b_\mu$ vanishes, and duality is restored, the electromagnetic force experienced by the magnetic current may be obtained from the former by a simple duality transformation. This raises an interesting question: once duality is lost in the present framework could the magnetic sources experience an electromagnetic force proportional to $b_\mu$?\\

Although simple, a precise answer to this question seems to be not trivial in this context where Lorentz symmetry is violated. Actually, in the presence of these both sorts of sources, the electromagnetic (density) force reads:
\begin{equation}
\partial_\mu\Theta^{\mu\nu}=\frac{d}{d\tau}p^\nu=F^{\nu\alpha}J_\alpha
+\left[ \tilde{F}^{\nu\alpha} + \frac{b^\nu A^\alpha}{2}\right]
k_\alpha\,,\label{divEMtensor}
\end{equation}
where $p^\mu$ is the energy-momentum density. Whether the extra term above is non-vanishing appears to be a non-trivial question. We have seem two possible hypothesis:

i) First, if the parameters $b_\mu$ do not affect the measurement of $A_\alpha$ and $k^\alpha$ (allow for good choice of inertial observers), then their scalar product is a genuine Lorentz invariant and identically vanishes. In this case, we get the usual Lorentz force acting on both electric and magnetic sources, and at this level things would happen as duality held.\\

ii) On the other hand, if $b_\mu$ are so that it intrinsically affects the potential and magnetic current measurements from the observers, then the quantity $A_\mu k^\mu$ is expected to gives a non-vanishing result. In this case, we would have an extra term in the electromagnetic force, which for point-like charge would get the form:
\begin{eqnarray}
\frac{d}{d\tau}P^\mu|_g=g\tilde{F}^{\mu\alpha}\,u_\alpha +
\frac{b^\mu}{2}\int\,A^\alpha k_\alpha d^3\vec{x}_\gamma\,,\label{FLorentzg}
\end{eqnarray}
where $k_\alpha$ is the magnetic current associated to $g\delta^3(\vec{x})$ and $d^3\vec{x}_\gamma=\gamma\,d^3\vec{x}$ the Lorentz invariant 3-volume element, so that the non-covariance of the force above is related to $b^\mu$-parameter in the last term. If the latter hypothesis is correct then the searching for Dirac monopole provides another way for detecting these possible violations. Notice also that $A^\mu{k}_\mu$ identically vanishes for the case of a static monopole. If Lorentz symmetry was no violated in this scenario, this quantity would always vanish and we would have a negative answer to the question raised above. However, Lorentz invariance cannot be invoked and a definite answer demands further investigation.\\
There is another involved issue related to the energy-momentum density (\ref{divEMtensor}) and consequently to the force on a magnetic pole, eq. (\ref{FLorentzg}). Note that in both expressions the potential appears. Although this is not so surprising, we should stress that once magnetic sources are concerned such a function carries a singular structure (a non-physical line along which the potential blows up). Therefore, in this background where $b_\mu$ is non-vanishing the energy-momentum density, $\Theta^{\mu\nu}$, automatically bears such a ill-definition. This is not expected to hold whenever integrating over the sources volume, in such a way that the extra term of the force (\ref{FLorentzg}) should be well-defined.\\

\section{Conclusions and Prospects}

\indent We have studied the issue of Dirac-like monopole in a (3+1)-dimensional Maxwell Electrodynamics supplemented by a Lorentz- and CPT-breaking term, $b_\mu\,A_\nu\tilde{F}^{\mu\nu}$, where $b_\mu$ a fixed 4-vector. A naive analysis of the equations of motion seems to be incompatible with magnetic sources. However, this inconsistency may be bypassed if we assume that the introduction of magnetic monopole is accompanied by the appearance of an induced electric current. We have realized that the current density is well-defined while its integration over all space leads to a blowing up result (at least for the case of a static monopole), so that these quantities are well-defined by means of a suitable cutoff or by assuming that $b_\mu$ is restricted to a finite region of space-time. We have also considered the dimensional reduction of the equations of motion to (2+1) dimensions, where a vectorial and a scalar models with Lorentz-breaking and magnetic sources were identified. The simplest case of a instanton-like pole breaking the Bianchi identity in the vectorial sector leads to the Henneaux-Teitelboim quantization condition of the mass parameter (the 3rd component survivor of $b_\mu$ from 4D). We also argue that a more general condition involving also $b_{\hat{\mu}}$ is expected to be obtained once we consider the coupling between both models which is performed by Lorentz-violating terms.\\

In addition, we have also considered the electromagnetic force acting on magnetic sources in this framework. Intimately connected to the Lorentz and CPT violations, we obtain an extra term proportional to $b_\mu$ in the force density, eq. (\ref{divEMtensor}), which involves the product $A_\mu k^\mu$. Whether this term is non-vanishing in this background appears to be a non-trivial issue concerning Dirac monopole in this context. We have raised two possibilities for such a point and emphasized that a definite answer demands further investigation.  

A question which is immediately brought about is whether such a similar
situation also takes place in the non-Abelian case. It would become even more
interesting and involved if we consider this issue in the framework of a
supersymmetric Abelian or Yang-Mills model, in the presence of the Lorentz-symmetry
violating term above. In such a case, the fermionic partner of the
background vector responsible for the breaking of Lorentz symmetry may be
taken with an electromagnetic charge, according to the structure of the matter supermultiplet of four-dimensional supersymmetry. Therefore, this charged background might,
in some special situation, compete with the charge induced by the magnetic
source, so that they cancel each other. The interplay between the charge
coming from the background dictated by supersymmetry and the charge induced by the magnetic current should be contemplated in some more detail. For that, it is crucial that the background vector be the four-dimensional gradient of the physical scalar which sits in the same superfield where the charged fermionic background appears. Finally, another interesting point to be investigated is the dimensional reduction
program from (4+1) to (3+1) dimensions in order to better understand the
role played by $b_\mu$-like term and how our quantization relation emerges
from a higher-dimensional point of view. In addition, the properties of the scalar (2+1)Electrodynamics, in the presence of both Lorentz-violating and magnetic sources terms, is under study and will appear elsewhere\cite{workinprogress}.\\
\vskip .8cm
\centerline{\bf \Large Acknowledgements\\}
\vskip .5cm \noindent The authors thank CAPES, CNPq, and FAPEMIG (Brazilian agencies) for the financial support.\\

\end{document}